\PassOptionsToPackage{unicode}{hyperref}
\PassOptionsToPackage{hyphens}{url}
\documentclass[
]{article}
\usepackage{amsmath,amssymb}
\usepackage{lmodern}
\usepackage{iftex}
\ifPDFTeX
  \usepackage[T1]{fontenc}
  \usepackage[utf8]{inputenc}
  \usepackage{authblk}
  \usepackage{textcomp} 
\else 
  \usepackage{unicode-math}
  \defaultfontfeatures{Scale=MatchLowercase}
  \defaultfontfeatures[\rmfamily]{Ligatures=TeX,Scale=1}
\fi
\IfFileExists{upquote.sty}{\usepackage{upquote}}{}
\IfFileExists{microtype.sty}{
  \usepackage[]{microtype}
  \UseMicrotypeSet[protrusion]{basicmath} 
}{}
\makeatletter
\@ifundefined{KOMAClassName}{
  \IfFileExists{parskip.sty}{%
    \usepackage{parskip}
  }{
    \setlength{\parindent}{0pt}
    \setlength{\parskip}{6pt plus 2pt minus 1pt}}
}{
  \KOMAoptions{parskip=half}}
\makeatother
\usepackage{xcolor}
\usepackage[margin=1in]{geometry}
\usepackage{graphicx}
\makeatletter
\def\maxwidth{\ifdim\Gin@nat@width>\linewidth\linewidth\else\Gin@nat@width\fi}
\def\maxheight{\ifdim\Gin@nat@height>\textheight\textheight\else\Gin@nat@height\fi}
\makeatother
\setkeys{Gin}{width=\maxwidth,height=\maxheight,keepaspectratio}
\makeatletter
\def\fps@figure{htbp}
\makeatother
\newlength{\cslhangindent}
\newlength{\csllabelwidth}
\newlength{\cslentryspacingunit} 
\newenvironment{CSLReferences}[2] 
 {
  \setlength{\parindent}{0pt}
  \ifodd #1
  \let\oldpar\par
  \def\par{\hangindent=\cslhangindent\oldpar}
  \fi
  \setlength{\parskip}{#2\cslentryspacingunit}
 }%
 {}
\usepackage{calc}

\ifLuaTeX
  \usepackage{selnolig}  
\fi
\IfFileExists{bookmark.sty}{\usepackage{bookmark}}{\usepackage{hyperref}}
\IfFileExists{xurl.sty}{\usepackage{xurl}}{} 
\hypersetup{
  pdftitle={spINAR: An R Package for Semiparametric and Parametric Estimation and Bootstrapping of Integer-Valued Autoregressive (INAR) Models},
  hidelinks,
  pdfcreator={LaTeX via pandoc}}

\title{spINAR: An R Package for Semiparametric and Parametric Estimation
and Bootstrapping of Integer-Valued Autoregressive (INAR) Models}
\author[1,*]{Maxime Faymonville}
\author[1]{Javiera Riffo}
\author[1]{Jonas Rieger}
\author[1]{Carsten Jentsch}
\affil[1]{Department of Statistics, TU Dortmund University}
\affil[*]{Corresponding author: Maxime Faymonville, faymonville@statistik.tu-dortmund.de}
\date{\today}

\begin{document}
\maketitle

\hypertarget{summary}{%
\section{Summary}\label{summary}}

While the statistical literature on continuous-valued time series
processes is vast and the toolbox for parametric, non-parametric and
semiparametric approaches is methodologically sound, the literature on
count data time series is considerably less developed. Such count data
time series models are usually categorized in parameter-driven and
observation-driven models. Among the observation-driven approaches, the
integer-valued autoregressive (INAR) models that rely on the famous
binomial thinning operation due to Steutel \& Van Harn (1979) are
arguably the most popular ones. They have a simple intuitive and easy
interpretable structure and have been widely applied in practice (Weiß,
2009). In particular, the INAR(\(p\)) model can be seen as the discrete
analogue of the well-known AR(\(p\)) model for continuous-valued time
series. The INAR(1) model was first introduced by Al-Osh \& Alzaid
(1987) and McKenzie (1985), and its extension to the INAR(\(p\)) model
by Du and Li (1991) is defined according to
\[X_t = \alpha_1 \circ X_{t-1} + \alpha_2 \circ X_{t-2} + \ldots + \alpha_p \circ X_{t-p} + \varepsilon_t, \]
with \(\varepsilon_t \overset{\text{i.i.d.}}{\sim} G\), where the
innovation distribution \(G\) has range
\(\mathbb{N}_0=\{0,1,2, \ldots\}\). The vector of INAR coefficients
\(\alpha = (\alpha_1, \ldots, \alpha_p)' \in (0,1)^p\) fulfills
\(\sum_{i=1}^p \alpha_i < 1\) and
\[\alpha_i \circ X_{t-i} = \sum\limits_{j=1}^{X_{t-i}} Z_j^{(t,i)}, \, Z_j^{(t,i)} \sim \text{Bin}(1, \alpha_i), \]
where ``\(\circ\)'' denotes the binomial thinning operator first
introduced by Steutel \& Van Harn (1979). Although many contributions
have been made during the last decades, most of the literature focuses
on parametric INAR models and estimation techniques. We want to
emphasize the efficient semiparametric estimation of INAR models (Drost,
Van den Akker, \& Werker, 2009).

\hypertarget{statement-of-need}{%
\section{Statement of need}\label{statement-of-need}}

INAR models find applications in a wide variety of fields such as
medical sciences, environmentology and economics. For example, Franke \&
Seligmann (1993) model epileptic seizure counts using an INAR(1) model,
Thyregod, Carstensen, Madsen, \& Arnbjerg-Nielsen (1999) use
integer-valued autoregressive models to model the dynamics of rainfall
and McCabe \& Martin (2005) to analyze wage loss claims data. They all
have in common assuming that the innovation distribution belongs to a
parametric class of distributions. Non- or semiparametric estimation of
the INAR model was not considered until Drost et al. (2009) came up with
their semiparametric estimation approach. A possible explanation is the
complexity of the semiparametric setup since despite in the AR case the
estimation in the INAR case cannot be based on the residuals: Even if
the autoregressive coefficients were known, observing the data does not
imply observing the innovations (Drost et al., 2009). Nonetheless, one
big advantage of semiparametric estimation is that we do not need to
make a parametric distribution assumption on the innovations. The
Poisson assumption is, for example, the most frequently used assumption
for innovations and is characterized by equidispersion. In most cases,
however, the data shows a higher variance than the mean value. The
question arises when the distance between these two moments is large
enough to not rather assume overdispersion, which would probably lead to
assume negative binomially or geometrically distributed innovations.
Furthermore, when dealing with low counts, we often observe many zeros
in the data. This could be a sign for a zero-inflated innovation
distribution such as the zero-inflated Poisson distribution (Jazi,
Jones, \& Lai, 2012). However, it is unclear at what point the zero is
represented frequently enough in the data set to justify such an
assumption. The mentioned points indicate that the assumption of an
appropriate innovation distribution is often critical, bearing in mind
that an incorrect assumption can lead to poor estimation performance.
With semiparametric estimation, we do not have to commit to an
innovation distribution, which makes this approach appealing.

To deal with count data time series,
\href{https://www.r-project.org/}{\texttt{R}} (R Core Team, 2023)
provides the package
\href{https://cran.r-project.org/web/packages/tscount/}{\texttt{tscount}}
(Liboschik, Fokianos, \& Fried, 2017) which, a.o., includes
likelihood-based estimation of parameter-driven count data time series
models which do not include INAR models and exclusively allows for
conditional Poisson or negative binomially distributed data. The R
package
\href{https://CRAN.R-project.org/package=ZINARp}{\texttt{ZINARp}}
(Medina Garay, de Lima Medina, \& Rossiter Araújo Monteiro, 2022) allows
to simulate and estimate INAR data by using MCMC algorithms for
estimation but the package is limited to parametric estimation of INAR
models, that is, of the INAR coefficients and of a parametrically
specified innovation distribution
\(\{G_\theta \, | \, \theta \in \mathbb{R}^q, \, q \in \mathbb{N}\}\)
where they only cover the cases of Poisson or zero-inflated Poisson
distributed innovations.The
\href{https://julialang.org/}{\texttt{Julia}} (Bezanson, Edelman,
Karpinski, \& Shah, 2017) package
\href{https://zenodo.org/record/7488440\#.Y9ky9ISZNaQ}{\texttt{CountTimeSeries}}
(Stapper, 2022) deals with integer counterparts of ARMA and GARCH models
and some generalizations including the INAR model. It covers the
parametric estimation setup for INAR models but does also not allow for
non-parametric estimation of the innovation distribution. Such a
semiparametric estimation technique that still relies on the binomial
thinning operation, but comes along without any parametric specification
of the innovation distribution was proposed and proven to be efficient
by Drost et al. (2009). Also neither of the three packages contains
procedures for bootstrapping INAR models within these parametric and
semiparametric setups. The \href{https://www.r-project.org/}{\texttt{R}}
package \href{https://github.com/MFaymon/spINAR}{\texttt{spINAR}} fills
this gap and combines simulation, estimation and bootstrapping of INAR
models in a single package. Both, the estimation and the bootstrapping,
are implemented semiparametrically and also parametrically. The package
covers INAR models of order \(p \in \{1,2\}\), which are mainly used in
applications.

\hypertarget{features}{%
\section{Features}\label{features}}

For the simulation of INAR data, our package allows for flexible
innovation distributions that can be inserted in form of a parametric
probability mass function (pmf) or by simply passing a user-defined
vector as pmf argument. Regarding the estimation, it allows for moment-
and maximum likelihood-based parametric estimation of INAR models with
Poisson, geometrically or negative binomially distributed innovations
(see for example Weiß (2018) for details), but the main contribution
lies in the semiparametric maximum likelihood estimation of INAR models
introduced by Drost et al. (2009) which they proved to be efficient.
Additionally, a finite sample refinement for the semiparametric setup
consisting of an estimation approach, that penalizes the roughness of
the innovation distribution as well as a validation function for the
penalization parameters is implemented (Faymonville, Jentsch, Weiß, \&
Aleksandrov, 2022). Furthermore, the package includes the possibility to
bootstrap INAR data. Again, the user is able to choose the parametric or
the more flexible semiparametric model specification and to perform the
(semi)parametric INAR bootstrap described in Jentsch \& Weiß (2017).

\hypertarget{acknowledgements}{%
\section{Acknowledgements}\label{acknowledgements}}

This research was funded by the Deutsche Forschungsgemeinschaft (DFG,
German Research Foundation) - Project number 437270842.

\hypertarget{references}{%
\section*{References}\label{references}}
\addcontentsline{toc}{section}{References}

\hypertarget{refs}{}
\begin{CSLReferences}{1}{0}
\leavevmode\vadjust pre{\hypertarget{ref-alosh}{}}%
Al-Osh, M. A., \& Alzaid, A. A. (1987). First-order integer-valued
autoregressive (INAR(1)) process. \emph{Journal of Time Series
Analysis}, \emph{8(3)}, 261--275.

\leavevmode\vadjust pre{\hypertarget{ref-julia}{}}%
Bezanson, J., Edelman, A., Karpinski, S., \& Shah, V. B. (2017). Julia:
A fresh approach to numerical computing. \emph{SIAM {R}eview},
\emph{59}(1), 65--98.
doi:\href{https://doi.org/10.1137/141000671}{10.1137/141000671}

\leavevmode\vadjust pre{\hypertarget{ref-drost}{}}%
Drost, F., Van den Akker, R., \& Werker, B. (2009). Efficient estimation
of auto-regression parameters and innovation distributions for
semiparametric integer-valued AR(\(p\)) models. \emph{Journal of the
Royal Statistical Society. Series B}, \emph{71, Part 2}, 467--485.

\leavevmode\vadjust pre{\hypertarget{ref-faym}{}}%
Faymonville, M., Jentsch, C., Weiß, C. H., \& Aleksandrov, B. (2022).
Semiparametric estimation of INAR models using roughness penalization.
\emph{Statistical Methods and Applications}.
doi:\href{https://doi.org/10.1007/s10260-022-00655-0}{10.1007/s10260-022-00655-0}

\leavevmode\vadjust pre{\hypertarget{ref-franke_med}{}}%
Franke, J., \& Seligmann, T. (1993). Conditional maximum-likelihood
estimates for INAR(1) processes and their applications to modeling
epileptic seizure counts. \emph{Developments in Time Series}, 310--330.

\leavevmode\vadjust pre{\hypertarget{ref-jazi}{}}%
Jazi, M., Jones, G., \& Lai, C. (2012). First-order integer valued AR
processes with zero inflated poisson innovations. \emph{Journal of Time
Series Analysis}, \emph{33}, 954--963.

\leavevmode\vadjust pre{\hypertarget{ref-jewe}{}}%
Jentsch, C., \& Weiß, C. H. (2017). Bootstrapping INAR models.
\emph{Bernoulli}, \emph{25(3)}, 2359--2408.

\leavevmode\vadjust pre{\hypertarget{ref-tscount}{}}%
Liboschik, T., Fokianos, K., \& Fried, R. (2017). {tscount}: An {R}
package for analysis of count time series following generalized linear
models. \emph{Journal of Statistical Software}, \emph{82}(5), 1--51.
doi:\href{https://doi.org/10.18637/jss.v082.i05}{10.18637/jss.v082.i05}

\leavevmode\vadjust pre{\hypertarget{ref-mc_wage}{}}%
McCabe, B., \& Martin, G. (2005). Bayesian predictions of low count time
series. \emph{International Journal of Forecasting}, \emph{21(2)},
315--330.

\leavevmode\vadjust pre{\hypertarget{ref-mck}{}}%
McKenzie, E. (1985). Some simple models for discrete variate time
series. \emph{Water Resources Bulletin}, \emph{21(4)}, 645--650.

\leavevmode\vadjust pre{\hypertarget{ref-zinarp}{}}%
Medina Garay, A. W., de Lima Medina, F., \& Rossiter Araújo Monteiro, T.
A. (2022). \emph{ZINARp: Simulate INAR/ZINAR(p) models and estimate its
parameters}. Retrieved from \\
\url{https://CRAN.R-project.org/package=ZINARp}

\leavevmode\vadjust pre{\hypertarget{ref-rcoreteam}{}}%
R Core Team. (2023). R: {A} {Language} and {Environment} for
{Statistical} {Computing}. R Foundation for Statistical Computing.
Retrieved from \url{https://www.R-project.org/}

\leavevmode\vadjust pre{\hypertarget{ref-manuel}{}}%
Stapper, M. (2022). ManuelStapper/CountTimeSeries.jl: v0.1.4.
doi:\href{https://doi.org/10.5281/zenodo.7488440}{10.5281/zenodo.7488440}

\leavevmode\vadjust pre{\hypertarget{ref-steutel}{}}%
Steutel, F. W., \& Van Harn, K. (1979). Discrete analogues of
self-decomposability and stability. \emph{Annals of Probability},
\emph{7(5)}, 893--899.

\leavevmode\vadjust pre{\hypertarget{ref-thy_rain}{}}%
Thyregod, P., Carstensen, J., Madsen, H., \& Arnbjerg-Nielsen, K.
(1999). Integer valued autoregressive models for tipping bucket rainfall
measurements. \emph{Environmetrics}, \emph{10}, 395--411.

\leavevmode\vadjust pre{\hypertarget{ref-dissweiss}{}}%
Weiß, C. H. (2009). \emph{Categorical times series analysis and
applications in statistical quality control}. dissertation.de.

\leavevmode\vadjust pre{\hypertarget{ref-bookweiss}{}}%
Weiß, C. H. (2018). \emph{An introduction to discrete-valued time
series} (1st ed.). Wiley.

\end{CSLReferences}

\end{document}